# Doping and dimensionality effects on the core-level spectra of layered ruthenates

Haizhong Guo,[1,2] Yi Li,[1] Darwin Urbina,[3] Biao Hu,[1] Rongying Jin,[1] Tijiang
Liu,[4] David Fobes,[4] Zhiqiang Mao,[4] E. W. Plummer,[1] and Jiandi Zhang[1,*]

[1]*Department of Physics and Astronomy, Louisiana State University, Baton Rouge, LA 70803, USA*
[2]*Beijing National Laboratory for Condensed Matter Physics, Institute of Physics,
Chinese Academy of Sciences, Beijing 100190, People's Republic of China*
[3]*Department of Physics, Florida International University, Miami, FL 33199, USA*
[4]*Physics Department, Tulane University, New Orleans, LA 70118, USA*
(Dated: October 26, 2018)

Core-level spectra of the $Mn$-doped $Sr_3Ru_2O_7$ and $Sr_{n+1}Ru_nO_{3n+1}$ ($n = 1, 2$ and $3$) crystals are investigated with X-ray photoelectron spectroscopy. Doping of $Mn$ to $Sr_3Ru_2O_7$ considerably affects the distribution of core-level spectral weight. The satellite of $Ru$ $3d$ core levels exhibits a substantial change with doping, indicating an enhanced electron localization across the doping-induced metal-insulator transition. However, the $Ru$ $3p$ core levels remain identical with $Mn$-doping, thus showing no sign of doping-induced multiple $Ru$ valences. In the $Sr_{n+1}Ru_nO_{3n+1}$ ($n = 1, 2$ and $3$), the $Ru$ $3d$ core-level spectra are similar, indicating that the chemical bonding environment around Ru ions remains the same for different layered compounds. While the $Sr$ $3d$ shallow core levels shift to higher binding energy with increasing $n$, suggesting their participation in $Sr$-O bonding with structural evolution.

## I INTRODUCTION

The layered ruthenates of the Ruddlesden-Popper series $Sr_{n+1}Ru_nO_{3n+1}$ (see Fig. 1), where $n$ ($n = 1, 2, 3$, ... $\infty$) is the number of layers of corner sharing $RuO_6$ octahedra per formula unit, display a remarkable array of complex electronic and magnetic properties.[1,2] The complexity, which is intimately related to the coexistence of competing nearly degenerate states which couple simultaneously active degrees of freedom: charge, lattice, orbital and spin states, is directly responsible for their tunability. Specifically, the properties of $Sr_{n+1}Ru_nO_{3n+1}$ exhibit strong dependence on the number ($n$) of $RuO_6$ octahedral layers in crystal structure, reflecting the effect of dimensionality in the system. Single-layered $Sr_2RuO_4$ ($n = 1$), as the most two-dimensional-like compound in the perovskite series, is an unconventional superconductor with possible spin-triplet pairing.[1,3–5] The bilayered $Sr_3Ru_2O_7$ ($n = 2$) shows behavior consistent with proximity to a metamagnetic quantum critical point.[6] The magnetic ground state of the triple-layer $Sr_4Ru_3O_{10}$ ($n = 3$) is poised between an itinerant metamagnetic and itinerant ferromagnetic state.[7–9] $SrRuO_3$ ($n = \infty$), regarded as a three-dimensional compound, is an itinerant ferromagnet with unusual transport characteristic.[10,11] On the other hand, the replacement of $Sr$ with $Ca$[12] or $Ru$ with other transition metal ions such as $Mn$[13,14] causes a metal-to-insulator transition (MIT), reflecting the manifestation on transport property by replacing ions with different sizes. Especially, the dilute Mn-doping provides a remarkably effective pathway of tuning on electronic structure beyond disorder-induced electron localization.[14] However, the nature of such a doping-induced MIT is still under investigation.

Photoelectron spectroscopy (XPS) has been widely used to study the chemical environment and the electronic structure of materials. In particular, the satellite structures in the core level photoemission spectra provide important information about the interactions of electrons in correlated systems such as transition-metal oxides (TMOs)[15]. One example is the XPS study of various ruthenates, where $Ru$ $3d$ core-level XPS spectra have satellites which are suggested as a result of two different screening mechanisms in the Mott-Hubbard picture for the MIT.[16] Since the MIT was observed with partial substitution of $Ru$ by $Mn$ in $Sr_3Ru_2O_7$,[13,14] one expects that the study of core level structure may shed light on the nature of doping-induced MIT in this system. Recent results of the X-ray absorption spectroscopy (XAS) taken from $Mn$-doped $Sr_3Ru_2O_7$ suggest that $Mn$ impurities do not exhibit the same 4+ valence as $Ru$, but act as 3+ acceptors and the observed MIT is purely electronic.[17] Therefore, it is expected that $Ru$ core-level XPS spectra may give further indication about doping-induced multiple $Ru$ valences ($Ru^{4+}$ and $Ru^{5+}$) and, more importantly, the nature of the doping-induced MIT. In addition, one expects that the change of the number $n$ of $RuO_6$ octahedral layers should affect the electronic structures and the electron correlation in the layered $Sr_{n+1}Ru_nO_{3n+1}$ series. However, the relationship between the changes in the electron correlation strength and the changes in the dimensionality ($n$) as well as the doping effect is far from clear. In this study, we used XPS to systematically investigate both the dimensionality and doping effects on the electronic structure and correlation in the $Sr_{n+1}Ru_nO_{3n+1}$ series ($n = 1, 2$ and $3$) and $Mn$-doped $Sr_3(Ru_{1-x}Mn_x)_2O_7$ system ($x = 0.0, 0.1$ and $0.2$).

## II EXPERIMENTS

The single crystals of $Sr_{n+1}Ru_nO_{3n+1}$ ($n = 1, 2$ and $3$) and $Mn$-doped $Sr_3(Ru_{1-x}Mn_x)_2O_7$ ($x = 0, 0.1,$ and 0.2) were grown by the floating zone technique.[8,9,13,18] The X-ray diffraction on these crystals did not reveal any impurity phase. The samples were cleaved at room temperature in ultrahigh vacuum (UHV) conditions to form a (001) surface and were immediately transferred into the $\mu$-metal-shielded analysis chamber equipped with a computer controlled low energy electron diffraction (LEED) diffractometer and XPS. The crystals were cleaved between two $SrO$ layers without breaking $RuO_6$ octahedra (See the cleavage planes in the Fig. 1). XPS measurements were carried out with a Phoibos-150 hemispherical energy analyzer (from SPECS) by using photons of energy of 1486.7 eV from a Micro-Focus 500 ellipsoidal crystal monochromator with Al $K\alpha$ X-ray source and focusing X-ray spot capabilities. A pass energy of 10 eV was used for the measurements. The overall energy resolution for the XPS spectra is 0.16 eV. The Fermi level ($E_F$) of a Au sample was measured and used to calibrate the binding energy of all core level spectra. All the XPS data were measured as the samples were at room temperature ($T = 300$ K). The base pressure of our system during the measurements was $2 \times 10^{-10}$ Torr.

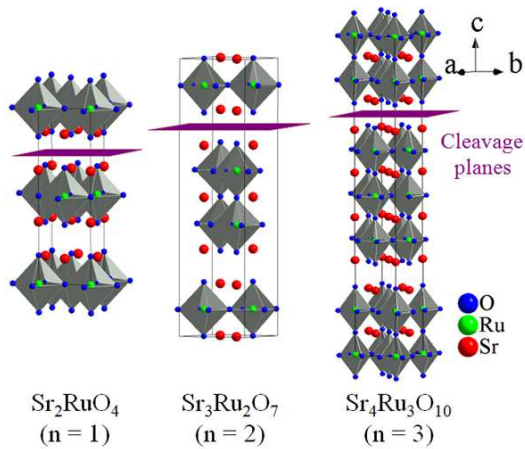

FIG. 1: (Color online) Units cells of $Sr_2RuO_4$ ($I4/mmm$), $Sr_3Ru_2O_7$ ($Pban$) and $Sr_4Ru_3O_{10}$ ($Pbam$). The different cites of oxygen ions in $Sr_3Ru_2O_7$ are labeled. The surfaces were created by cleaving the crystals between two $SrO$ layers without breaking $RuO_6$ octahedra.

The binding energy, linewidth and intensity of all measured core levels were determined through the standard fitting procedure. All core level peaks were fitted to Voigt function, which is a convolution of Lorentzian and Gaussian components. The Lorentzian component accounts for the intrinsic line profile of core level excitations with finite lifetime while the Gaussian one represents the contributions from other extrinsic broadening such as inhomogeneities in the surroundings of the emitting atoms as well as thermal and instrumental broadening. We used Shirley background profile[19] to fit the background of XPS spectra. The Shirley empirical function has been widely used to take account of the inelastic background of a core level peak with in a relatively small amount of inelastic scattering. This background profile is proportional to the integrated photoelectron intensity to higher kinetic energy.

## III RESULTS AND DISCUSSION

### A. Structures and LEED results

The structure of $Sr_2RuO_4$ is tetragonal with $I4/mmm$ (Fig. 1) symmetry while the structures of $Sr_3Ru_2O_7$ and $Sr_4Ru_3O_{10}$ are orthorhombic through rotations about the $c$-axis of the neighboring corner-sharing octahedra within each layer of the double or triple perovskite blocks.[20,21] These rotations result in a decrease of the $Ru-O-Ru$ angle in the $ab$ plane to $165^o$ in $Sr_3Ru_2O_7$ and $158^o$ in the middle layers of $Sr_4Ru_3O_{10}$ from the $180^o$ in tetragonal $Sr_2RuO_4$. Although the $RuO_6$ octahedra in these crystals have very similar size, the bond environment of cation $Sr$ changes with number $n$ of $RuO_6$ octahedral layers. Inside the crystals, $Sr$ has onlyone kind of bonding environment ($Sr-O(1)/Sr-O(3)$)in the single-layered $Sr_2RuO_4$[22] while two kinds of distinct bonding environments ($Sr-O(1)/Sr-O(3)$ and $Sr-O(2)/Sr-O(3)$) present in both the double[23]- and triple[20]-layered compounds (see Fig. 1).

Figure 2(a) shows a typical LEED pattern taken from the cleaved surface of $Sr_3Ru_2O_7$ at room temperature, showing excellent diffraction beams. In addition to the bright spots, there are weak spots (marked by arrows) related to the orthorhombic structure through the alternating in-plane rotation of $RuO_6$ octahedra about the (001) axis in bulk (see Fig. 2(b)). Our LEED images confirm that the surfaces of doped and undoped $Sr_3Ru_2O_7$ as well as $Sr_4Ru_3O_{10}$ have primary $p(1 \times 1)$ structure without reconstruction. However, a $(\sqrt{2} \times \sqrt{2})R45^o$ superstructure with respect to the tetragonal structure in the bulk was observed in $Sr_2RuO_4$, due to the rotation of the surface $RuO_6$ octahedra about the axis normal to the surface. This surface reconstruction corresponds to a bulk soft-phonon mode freezing into a static lattice distortion.[24]

### B. Doping dependence of core level spectra

Previous XAS[17] on $Sr_3(Ru_{1-x}Mn_x)_2O_7$ ($x = 0.0$ and 0.1) suggests that the driving mechanism for the observed doping-induced MIT is purely of electronic origin. To verify this, we have performed the XPS study on $Sr_3(Ru_{1-x}Mn_x)_2O_7$. Figure 3 presents the valence band and some characteristic core-level spectra of the $Sr_3(Ru_{1-x}Mn_x)_2O_7$ ($x = 0.0, 0.1,$ and $0.2$). With $Mn$-doping, a clear spectral weight transfer can be seen in the valence spectra (see Fig. 3(a)). The spectral weight of the quasi particle band considerably decreases and transfers to higher binding energy region of 3.0 to 10 eV. This should be associated with the enhanced electron localization with increasing $Mn$ doping and consequently the doping-induced MIT observed in the system. There is yet no clear microscopic understanding for the doping-induced MIT. Structural studies[14] indicate that when the system is doped with $Mn$ the $RuO_6$ octahedra are compressed along $c$-axis in a Jahn-Teller-like fashion, resulting in the bulking of octahedral layer. Such structural modifications with doping cause a change in the relative energy of the $t_{2g}$ orbitals, and consequently in the transport and magnetic properties.

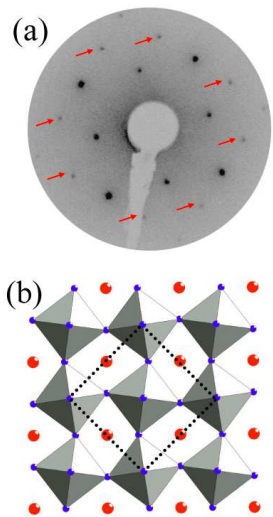

FIG. 2: (Color online) (a) A typical LEED pattern from a freshly cleaved $Sr_3Ru_2O_7$ (100) surface taken at room temperature with electron beam energy of 77 eV. Fractional spots are indicated by red arrows. (b) Schematic structure of a single $RuO_2$ plane. The surface unit cell is shown as a black dashed square. The red balls represent Sr atoms and blue ones represents O atoms. Ru ions are located in the center of the $RuO_6$ octahedra.

As shown in Fig. 3(b), (c) and (d), both $Sr$ and $Ru$ core levels shift to higher binding energy with increasing $Mn$ doping. In particular, $Sr$ and $Ru$ $3p$ core levels exhibit almost identical shifts with doping (see Fig. 3(e)). Such similar shifts may be attributed to final state effects due to the reduction of screening to core holes with $Mn$ doping. With the increase of $Mn$ doping, the valence electrons are becoming more localized and reduce their screening to core hole, thus resulting in an increase of apparent binding energy in the core level spectra. However, as we will discuss below, we also need to take into account the effects of doping-induced change of chemical environment known as initial state effects.

One important issue for the understanding the nature of $Mn$-doped $Sr_3Ru_2O_7$ is the valence of $Mn$ impurity in the doped compounds. In $Sr_3Ru_2O_7$, the valences of ions are $Sr^{2+}$, $Ru^{4+}$ and $O^{2-}$, respectively, which would suggest the substitution of $Ru^{4+}$ with $Mn^{4+}$ upon doping. However, the XAS measurements[17] on the $Sr_3(Ru_{0.9}Mn_{0.1})_2O_7$ suggested that $Mn$ impurities did not exhibit the same 4+ valence as $Ru$, but acted as 3+ acceptors. This conclusion was obtained by comparing the isotropic $Mn$ $L_{2,3}$-edge XAS data from $Sr_3(Ru_{0.9}Mn_{0.1})_2O_7$ with these from stoichiometric $Mn$ oxides of known valences such as $LaMnO_3$ and $Sr_3Mn_2O_7$. If the $Mn$ substituent in $Sr_3(Ru_{1-x}Mn_x)_2O_7$ acts as a $Mn^{3+}$ electron acceptor, a substantial charge transition disproportionation of $Mn^{4+} + Ru^{4+} \longrightarrow Mn^{3+} + Ru^{5+}$ should exist in the system, and the $Ru$ ions in doped samples should be mixed-valent with $Ru^{4+}/Ru^{5+}$. This would result in two components in the $Ru$ core level spectra associated with $Ru^{4+}$ and $Ru^{5+}$ ion, respectively. As long as these two components have difference in binding energy based on the scenario of initial state effect, one would expect that the lineshape of $Ru$ core spectra varies with doping. As shown in Fig. 3(f) where we shifted the $Ru$ $3p_{3/2}$ peaks measured from the $x = 0.1$ and 0.2 samples to lower binding energy in order to line up with the peak of the undoped compound. The $Ru$ $3p_{3/2}$ peaks of these three samples completely overlap, showing no change in lineshape with doping. Therefore, there is no indication of multiple $Ru$ valences induced by $Mn$-doping in the core spectra in contrast with the results of X-ray absorption spectroscopy. The doping-dependence of the $Mn$ $2p$ core-level spectra, with $2p_{1/2}$ at $\sim 653$ eV and $2p_{3/2}$ at $\sim 642$ eV, is shown in Fig. 4 (a) (see the inset). Because of the relatively weak $Mn$ XPS signal, we are unable to resolve any change of $Mn$ core levels with doping.

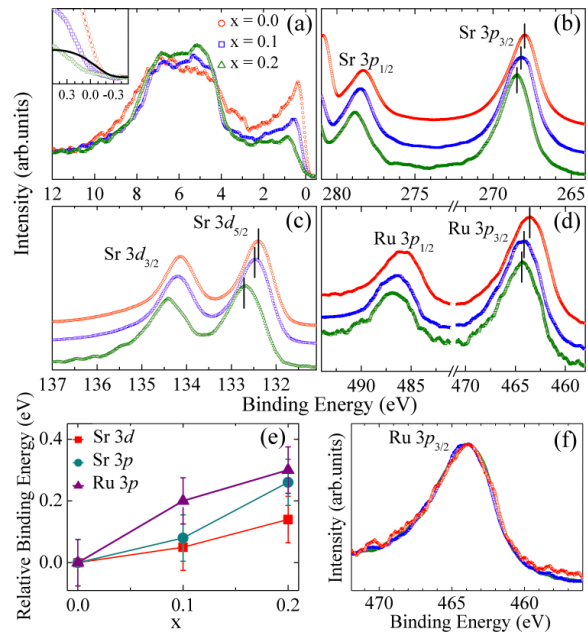

FIG. 3: (Color online) Valence and core-level photoemission spectra of the $Sr_3(Ru_{1-x}Mn_x)_2O_7$ ($x = 0.0$, 0.1, and 0.2) measured at $T = 300$ K: (a) Valence spectra; (b) $Sr$ $3p$; (c) $Sr$ $3d$; (d)$Ru$ $3p$ spectra, (e) Energy shifts of $Ru$ $3p$, $Sr$ $3p$ and $3d$ core level spectra relative to those in $Sr_3Ru_2O_7$ ($x = 0.0$) plotted against the doping level; and (f) The $Ru$ $3p_{3/2}$ core level after being shifted for the comparison of lineshape. The peak positions are marked by the solid bars. The zoom-in valence spectra close to $E_F$ compared with that taken from a Au sample (black solid curve) are shown in the inset of panel (a).

As we have mentioned above, $Mn$-doping induces a MIT in $Sr_3Ru_2O_7$[13,14] which should be reflected in the $Ru$ $3d$ spectral evolution. In fact, the $Ru$ $3d$ spectra exhibit strong doping-dependence. As shown in Fig. 4(a), both $Ru$ $3d_{3/2}$ and $3d_{5/2}$ core levels show double peaks, one labeled as "screened ($s$)" peak with lower binding energy and the other labeled as "unscreened ($u$)" peak with higher binding energy. For $Sr_3Ru_2O_7$, the $s$ peak of $Ru$ $3d_{5/2}$ is located at 280.76 eV and the $u$ one at 282.13 eV. The separation of these two peaks in both $Ru$ $3d_{3/2}$ and $3d_{5/2}$ core levels is about 1.47 eV for the undoped sample and 1.33 eV for both doped samples. These peaks in the $Ru$ $3d$ spectra shift to higher binding energy with doping. The spectral weight of the $s$ peak decreases while the $u$ peak increases with doping. The broad $u$ peak is the satellite (refereed to as shake-up) due to particle-hole excitations in the presence of the one-particle-hole state. Generally, the intensity of satellites is directly proportional to the degree of electron localization, thus providing a signature for the MIT in a system.

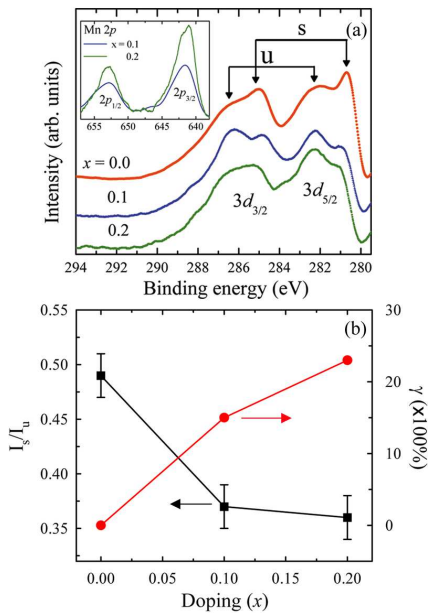

FIG. 4: (Color online) (a) Doping-dependence of $Ru$ $3d$ spectra with "screened ($s$)" and "unscreened ($u$)" components taken from $Mn$-doped $Sr_3Ru_2O_7$; (b) Intensity ratio of the fitted $s$ and $u$ peaks of the $Ru$ $3d_{5/2}$ and the relative change of the first moment of the intensity weighted average energy vs. $Mn$ doping (see details in text). The inset shows the doping dependence of $Mn$ $3p$ core level spectra.

To gain insight into the doping dependence, the spectra were analyzed through standard core level fitting by assuming that both $Ru$ $3d_{3/2}$ and $3d_{5/2}$ core-levels exhibit two-peak structure. Peak positions, widths, and intensities were determined by fitting with Voigt function after Shirley background subtraction. Peak fittings of the $Ru$ $3d_{3/2}$ and $d_{5/2}$ spectra into two-peak structure gave a reasonably good fit to the experimental data. The Lorentzian component, which represents the excitation spectra of core level states with finite lifetime, was found to be the dominant one for $s$ peaks. The fitting for $u$ peaks contains significant Gaussian broadening. The obtained relative intensity ratio of the $u$ and $s$ component of the $Ru$ $3d_{5/2}$ core level as a function of $Mn$ doping concentration is plotted in Fig. 4 (b). The relative intensity ratio systematically decreases as the $Mn$ concentration increases. This behavior is correlated with that of the spectral weight of the quasi-particle spectra in valence band with doping, suggesting the gradual enhancement of electron localization towards MIT. This behavior is also consistent with the results of the measured transport properties, including $Mn$ doping-induced MIT.[13,14].

In order to further understand the nature of these $Ru$ $3d$ core spectra, we have calculated the first-moment of the intensity weighted average binding energy vs. $Mn$-doping.[25,26] According to the basic sum rules for all excitations which generate both the main core level peak and the associated shakeup satellites in core level spectra, the first moment of the intensity-weighted energy distribution equals the single-particle Hartree-Fock eigenenergy, i.e.,

$$E_k^{HF} = \int_{-\infty}^{\infty} \epsilon A(\vec{k},\epsilon)d\epsilon \approx \sum_i \epsilon_i I_i / \sum_i I_i + Continuum \quad (1)$$

where $A(\vec{k},\epsilon)$ is the spectral function in the core level photoelectron excitations. For a core level spectrum with discrete main peak and satellites, the integration of the spectral function can be replaced by a simple summation of the intensity ($I_i$) weighted average energy, where index $i$ refers to different discrete peaks, and the excitation continuum. Changes in $E_k^{HF}$ truly reflect the chemical shifts related to the changes of bonding configurations caused (in this case) by doping. For $Sr_3(Ru_{1-x}Mn_x)_2O_7$, we can determine $E_k^{HF}$ of the $Ru$ $3d_{5/2}$ core level for different doping concentrations by taking into account both the $s$ and $u$ peaks. For simplicity, we neglect the possible change in the excitation continuum with doping. Figure 4(b) presents the doping-dependence of the relative shift $\gamma$ of $E_k^{HF}$ normalized to the energy difference $\Delta E$ between the $s$ and $u$ peak. i.e.,

$$\gamma(x) = [E_k^{HF}(x) - E_k^{HF}(x=0)]/\Delta E \quad (2)$$

where $\Delta E$ (= 1.33 eV) is the same for both $x = 0.1$ and 0.2 samples from our analysis. For the undoped compound, we have $\Delta E = 1.47$ eV, slightly larger than that found in the doped ones. It is shown [see Fig. 4(b)] that $\gamma$ of $E_k^{HF}$ does increase with increasing doping, thus indicating a chemical shift in the core level spectra. The increase of the binding energy indicates the enhanced localization of the wave functions of the initial states with doping. Therefore, the determination of $\gamma(x)$, which can be used as an order parameter, reveals the critical doping concentration ($x_c$) for the metal-to-insulator transition observed in the system.

### C. Dimensionality effect on core level spectra

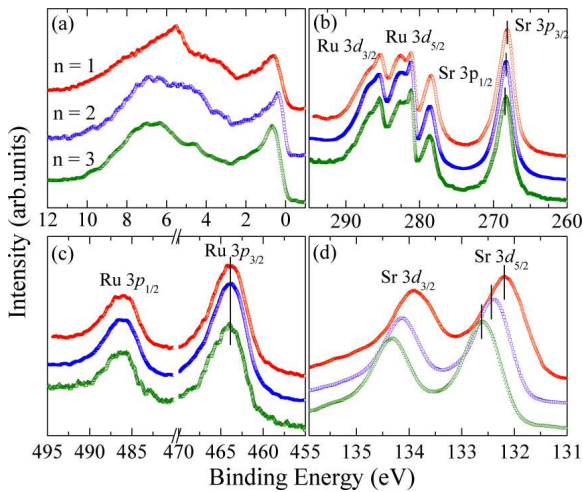

FIG. 5: (Color online) Valence and core-level photoemission spectra of the $Sr_{n+1}Ru_nO_{3n+1}$ ($n = 1, 2$ and $3$) measured at $T = 300$ K with monochromated Al $K\alpha$ X-ray source: (a) Valence spectra, (b) $Ru$ $3d$ and $Sr$ $3p$, (c) $Ru$ $3p$, and (d) $Sr$ $3d$ core level spectra. The peak positions are marked by the solid bars. The Fermi edge is calibrated with that of a Au sample.

Figure 5 presents the valence band and core-level spectra of the $Sr_{n+1}Ru_nO_{3n+1}$ ($n = 1, 2$ and $3$). The valence band spectra display a well defined Fermi edge for all three crystals. The $Ru$ $3d$ and $Sr$ $3p$ core levels are shown in Fig. 5(b). The positions, linewidth and relative intensity of these features in both valence and $Ru$ $3d$ exhibit similar change in spectral weight with different numbers of $RuO_6$ octahedra layers, $n$. The ratios of the coherent to incoherent part in the valence band and the ratio of the $s$ to $u$ component in the $Ru$ $3d$ core level increase slightly with $n$. The two ratios show a consistent behavior, indicating the electron correlation effects become smaller as the system gets closer to three dimensional. On the other hand, as shown in Fig. 5(c), the $Ru$ $3p_{1/2}$ (at 485.94.1 eV) and $3p_{3/2}$ (at 463.44 eV) core level spectra of $Sr_{n+1}Ru_nO_{3n+1}$ ($n = 1, 2$ and $3$) are almost identical. Tt is worth to notice that the surface of the $n = 1$ compound but not the $n = 2$ or $3$ one has a $(\sqrt{2} \times \sqrt{2})R45^o$ reconstruction. This may give rise to an additional change in the spectral weight of the $n = 1$ sample compared to the $n = 2$ or $3$ compound. Nevertheless, the similar behavior of the $Ru$ core level spectra is consistent with the similar rigid two-dimensional $RuO_6$ octahedra layer in all of these layered compounds. It would be interesting to study the similar core levels in $SrRuO_3$ (i.e., $n = \infty$) which is supposed to be three-dimensional rather than quasi two-dimensional. So far there is still no measurement of the core level spectra from $SrRuO_3$ single crystal available for comparison.

However, obvious changes have been observed in the $Sr$ core level spectra of the samples with different the layered number $n$ (see Fig. 5(b) and 5(d)), although their lineshape is unaltered. In order to compare these three compounds, we plote the shifts of $Sr$ $3p$ and $3d$ as well as $Ru$ $3p$ core spectra, as a function of the number of $RuO_6$ octahedra layers $n$ relative to $n = 1$ in Fig. 6. While $Ru$ core does not change, the $Sr$ cores exhibit increases of binding energy with $n$. Specifically, the $Sr$ $3d$ shallow core levels show stronger $n$-dependence while the shift of the $Sr$ $3p$ core is probably within the range of error bar. Interestingly, the linewidth of $n = 1$ sample ($\sim 0.77$ eV) is slightly larger than those from $n = 2$ and $3$ samples ($\sim 0.74$ eV). This may be attributed to larger surface broadening due to the surface reconstruction of $Sr_2RuO_4$.[24]

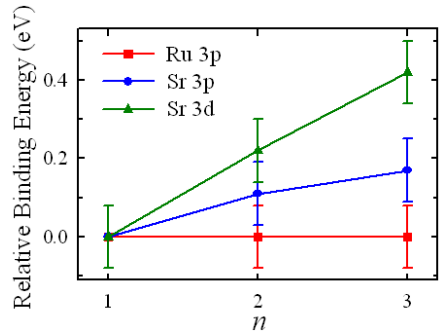

FIG. 6: (Color online) Measured energy shifts relative to the $Sr_2RuO_4$ ($n = 1$) plotted against the number of $RuO_6$ octahedra layers, $n$ in $Sr_{n+1}Ru_nO_{3n+1}$.

In order to understand the different shifts of the binding energy of $Sr$ core level spectra, it is necessary to examine the structural and bonding evolution of the cations with $n$ in ruthenate. As $n$ increases, the system evolves from quasi-two-dimensional to three dimensional structure (see Fig.3). Unlike the $Ru$ ions which are located inside $RuO_6$ octahedron, the $Sr$ ions has two distinct bonding environments in both $Sr_3Ru_2O_7$ and $Sr_4Ru_3O_{10}$. one bonding environment is the one with $Sr$-$O(1)$ bond along $c$-axis and $Sr$-$O(3)$ bond in $ab$ plane, referring to the $Sr$ ions in the cleavage plane (see Fig. 1). The other bonding environment is that of $Sr$-$O(2)$ bond along $c$-axis and $Sr$-$O(3)$ bond in $ab$ plane, referring to the $Sr$ ions within the $RuO_6$ layers. One can expect different chemical shifts due to the different bonding environments as well as the different bond lengths. In particular, the larger shift of $Sr$ $3d$ shallow core levels compared with that of $Sr$ $3p$ indicates a possible participation of $3d$ in bonding. Upon further increasing $n$, such shifts should gradually diminish as the system approaches a three-dimensional structure such as $SrRuO_3$ (i.e., $n = \infty$).

## IV SUMMARY

In summary, we have investigated the core-level XPS spectra of the layered ruthenates of the $Sr_{n+1}Ru_nO_{3n+1}$ ($n = 1, 2$ and $3$) and $Mn$-doped $Sr_3Ru_2O_7$. We observed that the dimensionality does not affect the $Ru$ but $Sr$ core levels. The $Ru$ $3d$ core-level spectra maintain similar structure for crystals with different octahedral layers

($n$), suggesting that the chemical bonding environment around Ru ions remains the same for different layered compounds. While the $Sr$ $3d$ shallow core levels shift to higher binding energy with increasing $n$, indicating a variation of $Sr$-O bonding with structural evolution. On the other hand, the core levels of $Sr_3(Ru_{1-x}Mn_x)_2O_7$ show strong dependence of doping concentration ($x$), revealing the enhanced electron localization with doping toward to MIT. However, the lineshape and linewidth of the $Ru$ $3p$ core levels remain identical with doping, showing no sign of multiple $Ru$ oxidation states in $Mn$-doped system.

## ACKNOWLEDGMENTS

This work was supported by US National Science Foundation Grant No. DMR-0346826. The XPS setup was obtained through the support from US DOD Grat No. W911NF-07-1-0532. Work at Tulane is supported by DOE under DE-FG02-07ER46358 (personnel support), NSF under grant DMR-0645305 (support for equipment), the DOD ARO under Grant No. W911NF-08-C0131 (support for materials), and Research Corporation.